\documentstyle[preprint,aps]{revtex}
\begin{document}
\preprint{ SPIN-1999/15, THU-99/12}
\title{Holography and Rotating AdS Black Holes}
\author{David S. Berman$^1$\footnote{d.berman@phys.uu.nl} and
Maulik K. Parikh$^2$\footnote{m.parikh@phys.uu.nl} }
\address{${}^1$Institute for Theoretical Physics, University of Utrecht,\\ 
Princetonplein 5, 3584 CC Utrecht, The Netherlands \\
${}^2$Spinoza Institute, University of Utrecht,\\
P. O. Box 80 195, 3508 TD Utrecht, The Netherlands}

\maketitle
\begin{abstract}

We probe the AdS/CFT correspondence by comparing the thermodynamics of a
rotating black hole in five-dimensional anti-de Sitter space with that of a
conformal field theory on $S^3$, whose parameters come from the boundary
of spacetime. In the high temperature limit, we find agreement 
between gauge theory and 
gravity in all thermodynamic quantities up to the same factor of 4/3 
that appears for nonrotating holes.


\end{abstract}

\def	\beq	{\begin{equation}}
\def	\eeq	{\end{equation}}
\def	\lf	{\left (}
\def	\rt	{\right )}
\def	\a	{\alpha}
\def	\lm	{\lambda}
\def	\D	{\Delta}
\def	\r	{\rho}
\def	\th	{\theta}
\def	\rg	{\sqrt{g}}
\def	\Slash	{\, / \! \! \! \!}
\def	\comma	{\; , \; \;}
\def	\Li	{\rm Li_4}
\def 	\pl 	{\partial}
\def 	\del 	{\nabla}

\section{Introduction}

One of the hallmarks of the duality revolution in string theory has been
the linking of apparently unrelated areas in physics via unexpected pathways. 
The AdS/CFT correspondence \cite{adscft}
is a striking example of this; by analyzing Dirichlet three-branes, 
it connects gravity in a particular
background to a strongly coupled gauge theory. More specifically, the
correspondence says that
IIB string theory in a background of five-dimensional anti-de Sitter space 
times a five-sphere is dual to the large N limit of ${\cal N} = 4$ 
supersymmetric Yang-Mills theory in four dimensions.
(See \cite{adsreview} for an extensive review of this vast subject.)

{}From a supergravity point of view, a 
Dirichlet p-brane is simply a charged extended black hole. Moreover, the
string theory description of D-branes allows one to determine its
world-volume action; to lowest order in $\a '$ this is super Yang-Mills.
In earlier work \cite{fourthirds}, 
the entropy of the black brane solution was correctly given to within 
a numerical coefficient by the entropy of the field theory on the brane.
Later, this match was extended to the case of rotating branes
\cite{roberto1,roberto2,steve,steve2,steve3}.

The AdS/CFT correspondence extends the relationship between gauge theory
and gravity from providing a description of a particular brane solution
to describing the physics of the entire supergravity background by a
dual conformal field theory in one dimension less. As such this
realizes the principle of holography \cite{hologerard,hololenny},
the notion that the physics of the bulk is imprinted on its boundary.

Black holes provide an arena in which this correspondence 
between gravity and gauge theory may be examined. 
For nonrotating AdS black holes \cite{hawkingpage}, the thermodynamics
has been described by thermal conformal field theory
\cite{holographyads}. Recently a five-dimensional 
rotating black hole embedded in anti-de
Sitter space has been discovered \cite{5dadskerr}.
Since rotation introduces an extra dimensionful parameter,
the conformal field theory entropy is not so tightly constrained by the 
combination of extensivity and dimensional analysis; a successful
correspondence between thermodynamic quantities is much more nontrivial.
Our purpose in this paper, therefore,
is to probe the correspondence by extracting the thermodynamics 
of the new rotating black hole from a dual conformal field 
theory in four dimensions.

We begin by demonstrating
the holographic nature of the duality for nonrotating black holes: the
thermodynamics of a nonrotating black hole in anti-de Sitter space
emerges from a thermal conformal field theory whose thermodynamic
variables are read off from the boundary of the black hole spacetime.
In the high temperature limit, the field theory calculation gives the
correct entropy of the Hawking-Page black hole up to a factor of 4/3.

We then describe the new rotating Kerr-AdS black
hole solution and show how its thermodynamic properties
can be recovered from the dual field theory, in the high temperature limit.
In that limit, the entropy, energy and angular momentum, as derived
from the statistical mechanics of the field theory, all agree with
their gravitational counterparts, again up to a common factor of 4/3.

\section{AdS/CFT Correspondence for Nonrotating Holes}

The five-dimensional Einstein-Hilbert action with a cosmological
constant is given by 
\beq 
I = - {1 \over 16 \pi G_5} \int d^5 x \sqrt{-g} \lf R + 12 l^2 \rt \; , 
\eeq 
where $G_5$ is the five-dimensional Newton
constant, $R$ is the Ricci scalar, the cosmological constant is
$\Lambda = - 6 l^2$, and we have neglected a surface term at infinity.
Anti-de Sitter solutions derived from this action can be embedded in
ten-dimensional IIB supergravity such that the supergravity background is of
the form $AdS_5 \times S^5$. The AdS/CFT correspondence then states that there 
is a dual conformal field theory in four dimensions from which one can 
extract the physics.

The line element of a ``Schwarzschild'' black hole in anti-de Sitter
space \cite{hawkingpage} in five spacetime dimensions can be written as 
\beq 
ds^2 = - \lf 1 - {2 M G_5 \over r^2} + r^2 l^2 \rt dt^2 +  
\lf 1 - {2 M G_5 \over r^2} + r^2 l^2 \rt^{\! \! -1} dr^2 
+ r^2 d \Omega^2_3 \; .  \label{HPds2}
\eeq
This solution has a horizon at $r = r_+$ where 
\beq 
r_+^2 = {1 \over 2 l^2} \lf -1  + \sqrt{1 + 8M G_5 \, l^2} \rt \; . \label{r+} 
\eeq 
The substitution $\tau = i t$  makes the metric positive definite and, by
the usual removal of the conical singularity at $r_+$, yields a
periodicity in $\tau$ of 
\beq 
\beta = {2 \pi r_+ \over 1 + 2r_+^2 l^2 } \; ,\label{invtemp}
\eeq 
which is identified with the inverse temperature
of the black hole. The entropy is given by  
\beq 
S = {A \over 4 G_5} = {\pi^2 r_+^3 \over 2 G_5} \; , \label{S}  
\eeq 
where $A$ is the ``area'' (that is 3-volume) of the horizon.

We shall take the dual conformal field theory to be ${\cal N} = 4$,
U(N) super-Yang-Mills theory.  
But since it is only possible to do calculations in the weak coupling
regime, we shall consider only the free field limit
of Yang-Mills theory. Then, in the high-energy regime which 
dominates the state counting, the spectrum of free fields on a sphere 
is essentially that of blackbody radiation in flat space, with
$8N^2$ bosonic and $8N^2$ fermionic degrees of freedom. The
entropy is therefore
\beq 
S_{\rm CFT} =
 {2 \over 3} \pi^2 N^2 \, V_{\rm CFT} \, T^3_{\rm CFT} \; . \label{ent} 
\eeq
We would like to evaluate this ``holographically", i.e. by substituting
physical data taken from the boundary of the black hole spacetime. At fixed
$r \equiv r_0 \gg r_+$, the boundary line element tends to 
\beq 
ds^2 \to r_0^2 \left [ - l^2 dt^2 + d \Omega_3^2 \right ] \; .
\eeq
The physical temperature at the boundary is consequently red-shifted to
\beq 
T_{\rm CFT} = {T_{BH} \over \sqrt{-g_{tt}}} = {T_{BH} \over lr_0} \; ,
\eeq 
while the volume is
\beq 
V_{\rm CFT} = 2 \pi^2 r_0^3 \; .  
\eeq
To obtain an expression for $N$, we invoke the AdS/CFT correspondence.
Originating in the near
horizon geometry of the D3-brane solution in IIB supergravity,
the correspondence \cite{adscft}, relates $N$ to the
radius of $S^5$ and the cosmological constant:
\beq 
R^2_{S^5} = \sqrt{4 \pi g_s {\a '}^2 N} = {1 \over l^2} \; .
\eeq 
Then, since
\beq 
(2 \pi)^7 g_s^2 {\a '}^4 = 16 \pi G_{10} =  16 {\pi^4 \over l^5} G_5 \; ,
\eeq 
we have
\beq 
N^2 = {\pi \over 2 l^3 G_5} \; . \label{N}
\eeq 
Substituting the expressions for $N$, $V_{\rm CFT}$ and $T_{\rm CFT}$ into 
Eq. (\ref{ent}), we obtain
\beq
S_{\rm CFT} = {1 \over 12} {\pi^2 \over l^6 G_5} \lf {1 + 2r_+^2 l^2
\over r_+} \rt ^{\! \! 3}  \; ,
\eeq 
which, in the high temperature limit $r_+ l \gg 1$, reduces to 
\beq
S_{\rm CFT}  = {2 \over 3} {\pi^2 r_+^3 \over G_5} = {4 \over 3}
S_{\rm BH}\; , 
\eeq 
in agreement with the black hole result,
Eq. (\ref{S}), but for a numerical factor of 4/3.

Similarly, the red-shifted energy of the conformal field theory matches the
black hole mass, modulo a coefficient.
The mass above the anti-de Sitter background is 
\beq 
M' = {3 \pi \over 4} M \; .
\eeq 
This is the AdS equivalent of the ADM mass, or energy-at-infinity. 
The corresponding expression in the field theory is
\beq 
U^{\infty}_{\rm CFT} = \sqrt{-g_{tt}} 
{\pi^2 \over 2} N^2 \, V_{\rm CFT} \, T^4_{\rm CFT}
= {\pi \over 2} r_+^4 l^2 = {4 \over 3} M' \; , \label{U}
\eeq 
where $U^{\infty}_{\rm CFT}$ is the conformal field theory energy
red-shifted to infinity, and we have again taken the $r_+ l \gg 1$ limit.
The $4/3$ discrepancy in Eqs. (\ref{S}) and (\ref{U}) is construed to be
an artifact of having calculated the gauge theory entropy in the free
field limit rather than in the strong coupling limit required by the
correspondence; intuitively, one expects the free energy to decrease
when the coupling increases. The 4/3 factor was first noticed in the context 
of D3-brane thermodynamics \cite{fourthirds}. 
Our approach differs in that we take
the idea of holography at face value, by explicitly reading physical
data from the boundary of spacetime; nonetheless, Eq. (\ref{N}) refers
to an underlying brane solution.

At this level, the correspondence only goes
through in the high temperature limit. Since the only two scales in the 
thermal conformal field theory are $r_0$ and $T_{\rm CFT}$, high temperature 
means that $T_{\rm CFT} \gg 1 / r_0$, allowing us to neglect 
finite-size effects.

\section{Five-Dimensional Rotating AdS Black Holes}

The general rotating black hole in five dimensions has two independent
angular momenta. Here we consider the case of a rotating black hole 
with one angular momentum in an ambient AdS space. 
The line element is \cite{5dadskerr}
\begin{eqnarray}
ds^2 & = & - {\D \over \r^2} \lf dt - {a \sin^2 \th \over \Xi _a} d \phi 
\rt ^{\! \! 2} + {\D_\th \sin^2 \th \over \r^2} \lf a \, dt - {\lf r^2 + a^2
\rt \over \Xi} d\phi \rt ^{\! \! 2} \nonumber \\ & & + {\r^2 \over \D} dr^2 +
{\r^2 \over \D_\th} \, d \th^2  + r^2 \cos^2 \th \, d \psi ^2 \; ,
\label{ds2}
\end{eqnarray}
where $0 \leq \phi , \psi \leq 2 \pi$ and $0 \leq \th \leq \pi / 2$, and
\begin{eqnarray}
\D & = & \lf r^2 + a^2 \rt \lf 1 + r^2 l^2 \rt - 2MG_5 \nonumber \\
\D_\th & = & 1 - a^2 l^2 \cos^2 \th \nonumber \\ \r^2 & = & r^2 + a^2
\cos^2 \th \nonumber \\ \Xi & = & 1 - a^2 l^2 \; .
\end{eqnarray}
This solution is an anti-de Sitter space with curvature given by 
\beq 
R_{ab} = - 4 l^2 g_{ab} \; .
\eeq 
The horizon is at 
\beq 
r^2_+ = {1 \over 2 l^2} \lf - (1 + a^2 l^2) + 
\sqrt{(1 + a^2 l^2)^2 + 8 M G_5 \, l^2} \rt \; ,
\eeq
which can be inverted to give
\beq 
M G_5 = {1 \over 2}(r_+^2 + a^2)(1+r_+^2l^2) \; .
\eeq
The entropy is one-fourth the ``area'' of the horizon:
\beq 
S = {1 \over 2 G_5} {\pi^2 \lf r_+^2 +a^2\rt r_+ \over \lf 1-a^2l^2\rt} \; . \label{SBH} 
\eeq 
The entropy diverges in two different limits: $r_+ \to \infty$ and $a^2
l^2 \to 1$. The first of these descibes an infinite temperature and
infinite radius black hole, while the second corresponds to 
``critical angular velocity'', at which the Einstein universe at
infinity has to rotate at the speed of light. The
inverse Hawking temperature is 
\beq 
\beta = {2 \pi \lf r_+^2 + a^2 \rt
\over r_+ \lf 1 + a^2 l^2 + 2 r_+^2 l^2 \rt} \; . \label{beta} 
\eeq 
The mass above the anti-de Sitter background is now 
\beq 
M' = {3 \pi \over 4 \Xi} M  \; , \label{mass} 
\eeq
the angular velocity at the horizon is
\beq
\Omega_H = {a \Xi \over r_+^2 + a^2} \; ,
\eeq
and the angular momentum is defined as
\beq  
J_{\phi}=  {1 \over 16 \pi}
\int_S \, \epsilon_{abcde} \del^{d} \psi^{e} dS^{abc}
= {\pi M a \over {2 \Xi^2}} \; , \label{jbh}
\eeq  
where $\psi^a= \lf {\pl \over \pl \phi} \rt ^{\! a}$ 
is the Killing vector conjugate to the angular momentum
in the $\phi$ direction, and S is the boundary of a hypersurface normal to
$\lf {\pl \over \pl t} \rt ^{\! a}$, with $dS^{abc}$ being the volume
element on S.

Following methods discussed in \cite{hawkingpage,thermalphase}, one
can derive a finite action for this solution from the regularized spacetime 
volume after an appropriate matching of hypersurfaces at large $r$.
The result is
\beq 
I = {\pi^2 \lf r_+^2 + a^2 \rt ^{\! \! 2} (1 - r_+^2 l^2) \over 4 G_5 \Xi r_+
(1 + a^2 l^2 + 2 r_+^2 l^2 )} \; . \label{f} 
\eeq
As noted in \cite{thermalphase}, 
the action changes sign at $r_+ l = 1$, signalling the
presence of a phase transition in the conformal field theory. For $r_+ l > 1$,
the theory is in an unconfined phase and has a free energy proportional to
$N^2$. One can also check that the action satisfies the thermodynamic relation 
\beq  
S = \beta(M' - J_{\phi} \Omega _H ) - I \; .
\eeq 
It is interesting to note that, by formally dividing both the free energy, 
$F = I / \beta$, and the mass by an arbitrary volume, one obtains an
equation of state: 
\beq 
p = {1 \over 3} {r_+^2 l^2 - 1 \over r_+^2 l^2 + 1} \, \rho \; , \label{es} 
\eeq 
where $p = - F/V$ is the pressure, and $\rho$ is the energy density. 
In the limit $r_+ l \gg 1$ that we have been taking, this equation becomes 
\beq  
p = {1 \over 3} \rho  \; , 
\eeq 
as is appropriate for the equation of state of a conformal
theory. This suggests that if a conformal field theory is to
reproduce the thermodynamic properties of this gravitational solution,
it has to be in such a limit.

\section{The dual CFT description}

The gauge theory dual to supergravity on $AdS_5 \times S^5$ is ${\cal
N} = 4$ super Yang-Mills with gauge group $U(N)$ where $N$ tends to
infinity \cite{adscft}.
The action is 
\beq 
S = \int d^4 x \rg ~ {\rm Tr}  \lf  -{1 \over 4 g^2}
F^2 + {1 \over 2} \lf D \Phi \rt ^2 + {1 \over 12} R \Phi^2 +
\bar{\psi} \Slash D \psi \rt \; .
\eeq 
All fields take values in the
adjoint representation of U(N). The six scalars, $\Phi$,
transform under $SO(6)$ R-symmetry, while the four Weyl fermions, $\psi$,
transform under $SU(4)$, the spin cover of $SO(6)$.  The scalars are
conformally coupled; otherwise, all fields are massless. We shall again
take the free field limit.
The angular momentum operators can be computed from the relevant components of
the stress energy tensor in spherical coordinates. 
This approach is to be contrasted 
with \cite{steve,steve2,steve3,finnmirjam} 
in which generators of R-rotations
are used corresponding to spinning D-branes.

The free energy of the gauge theory is given by
\beq 
F_{\rm CFT} = +T_{\rm CFT} \sum_i \eta_i \int_0^{\infty} dl_i 
\int d m^{\phi}_i \int
d m^{\psi}_i  \ln \lf 1 - \eta_i e^{- \beta (\omega _i - m^{\phi}_i
\Omega_{\phi})} \rt \; , \label{free}
\eeq 
where $i$ labels the particle species, $\eta = + 1$
for bosons and -1 for fermions, $l_i$ is the quantum number
associated with the total orbital angular momentum of the ith particle, and
$m^{\phi ( \psi )}_i$ is its angular momentum component in the 
$\phi (\psi)$ direction. Here $\Omega$ plays the role of a ``voltage''
while the ``chemical potential'' $m^{\phi} \Omega$ serves to constrain
the total angular momentum of the system.

The free energy is easiest to evaluate in a corotating frame, which
corresponds to the constant-time foliation choice of 
hypersurfaces orthogonal to $t^a$. Since, at constant $r = r_0$, the
boundary has the metric
\begin{eqnarray}
ds^2 & = & r_0^2 \left [ - l^2 dt^2 + {2a l^2 \sin^2 \th \over \Xi} dt \,
d\phi + {\sin^2 \th \over \Xi} d \phi^2 + {d\th^2 \over \D_\th} +
\cos^2 \th \, d\psi^2 \right ] \; ,
\end{eqnarray}
the constant-time slices of the corotating frame have a spatial volume of
\beq 
V =  {2 \pi^2 r_0^3 \over{ 1-a^2l^2}}\; . \label{volumecft}
\eeq
The spectrum of a conformally coupled field on $S^3$ is essentially given by
\beq
\omega _l \sim {l \over r_0} \; ,
\eeq
where $l$ is the quantum number for total orbital angular momentum.
Eq. (\ref{free}) can now be evaluated by making use of the identities
\begin{eqnarray}
\int_0^{\infty} dx \, x^n \ln \lf 1 - e^{-x + c} \rt  & = & - \Gamma(n+1)
{\rm Li}_{n+2} (e^c) = - \Gamma(n+1) \sum_{k = 1}^{k = \infty}  {e^{kc}
\over k^{n+2}} \nonumber \; , 
\\ \int dx \, x \, {\rm Li}_2(e^{-ax+c}) &=& - {1 \over
a^2} \left [ ax \, {\rm Li}_3(e^{-ax+c}) + {\rm Li}_4(e^{-ax+c}) \right ] \; ,
\end{eqnarray}
where ${\rm Li}_n$ is the nth polylogarithmic function, defined by the
sum above. The result is
\beq 
F_{\rm CFT} = -{\pi^4 \over 24} {r_0 \over {1 \over r_0^2}
- \Omega^2} \, (8 N^2) \, {T^4_{\rm CFT}} \; ,
\eeq
yielding an entropy of 
\beq
S_{\rm CFT} = {2 \over 3} {\pi^5 \over l^3 G_5} {r_0^3 \over 1 -
\Omega ^2 r_0^2} \, T_{\rm CFT}^3 \; .	\label{entropycft} 
\eeq 
The physical temperature that enters the conformal field theory is
\beq 
T_{\rm CFT} = {1 \over l r_0} T_{\rm BH} \; . \label{tempcft} 
\eeq 
Similarly, the angular velocity is scaled to 
\beq 
\Omega_{\rm CFT} = {al^2 \over l r_0} \; . \label{omegacft} 
\eeq 
Substituting Eqs. (\ref{tempcft}) and (\ref{omegacft}) into
Eq. (\ref{entropycft}) and taking the
high temperature limit as before, we have
\beq 
S_{\rm CFT} = {2 \over 3 G_5} {\pi^2 r_+^3 \over (1 - a^2 l^2)}  = {4 \over 3}
S_{\rm BH} \; .  
\eeq 
The inclusion of rotation evidently does not
affect the ratio of the black hole and field theory entropies.

In the corotating frame, the free energy is simply of the form $N^2 V T^4$,
with the volume given by Eq. (\ref{volumecft}). However, with respect to
a nonrotating AdS space, the free energy takes a more complicated form
since now the volume is simply $2 \pi^2 r_0^3$. By keeping this volume and
the temperature fixed, one may calculate the angular momentum of the system
with respect to the nonrotating background:
\beq 
J^{\rm CFT}_{\phi} = \left. - {\pl
F \over \pl \Omega} \right |_ {V,T_{\rm CFT}} 
= { a r_+^4 \pi \lf 1 + a^2 l^2 +2r_+^2 l^2 \rt ^4 \over 
48 l^6 \Xi^2 \lf r_+^2 + a^2 \rt ^4} \; .
\eeq
In the usual $r_+ l \gg 1$ limit, we obtain
\beq  
J_{\phi}^{\rm CFT} = {2 \pi M a \over 3 \Xi^2} = {4 \over
3} J_{\phi}^{\rm BH} \; ,
\eeq
so that the gauge theory angular momentum is proportional to the black
hole angular momentum, Eq. (\ref{jbh}), with a factor of 4/3.

The black hole mass formula, Eq. (\ref{mass}), refers to the energy above
the nonrotating anti-de Sitter background. We should therefore compare 
this quantity with the red-shifted energy in the conformal field theory.
Here a slight subtlety enters. 
Since the statistical mechanical calculation gives the energy in the
corotating frame, we must add the center-of-mass rotational energy
before comparing with the black hole mass. Then we find that
\beq
U^{\infty}_{\rm CFT} = \sqrt{-g_{tt}} \lf U_{\rm corotating} + J_{\rm CFT} 
\Omega_{\rm CFT} \rt = {4 \over 3} M' \; ,
\eeq
with $M'$ given by Eq. (\ref{mass}), evaluated at high temperature. Using
$U^{\infty}_{\rm CFT} = \sqrt{-g_{tt}} U_{\rm CFT}$ and previous expressions
for thermodynamic quantities, one may check that the first law 
of thermodynamics is satisfied.

\section{Discussion}

There are two interesting aspects of these results.
The first is that the same relative factor that appears in the entropy
appears in the angular momentum and the energy. A priori, one has no
reason to believe that the functional form of the free energy will be
such as to guarantee this result (see, for example,\cite{steve}). 
The second is that the relative factor of 4/3 in the entropy is 
unaffected by rotation. Indeed, one could
expand the entropy of the rotating system in powers and inverse
powers of the 't Hooft coupling. The correspondence implies that 
\beq 
S_{\rm CFT} = \sum_m a_m \lm^m = \sum_n b_n \lf {1 \over \sqrt{\lm}} \rt
^{\! \! n} = S_{\rm BH} \; .
\eeq 
We may approximate the series on the gauge theory side as
$a_0$ and on the gravity side as $b_0$. Then, generically, we would expect
these coefficients to be functions of the dimensionless rotational parameter
$\Xi$ so that $a_0 (\Xi) = f(\Xi) b_0(\Xi)$ with $f(\Xi = 1) =
4/3$. Our somewhat unexpected result is that $f(\Xi) = 4/3$ has, in
fact, no dependence on $\Xi$.

\section{Acknowledgments}

We would like to thank Jos\'e Barb\'on, Roberto Emparan, and 
Kostas Skenderis for helpful discussions. D. B. is supported by
European Commission TMR programme ERBFMRX-CT96-0045.
M. P. is supported by the Netherlands
Organization for Scientific Research (NWO).


\begin{thebibliography}{99}

\bibitem{adscft} J. Maldacena, ``The Large N Limit of Superconformal
field theories and supergravity,'' Adv. Theor. Math. Phys. {\bf 2} (1998)
231, hep-th/9711200.

\bibitem{adsreview} O. Aharony, S. S. Gubser, J. Maldacena, H. Ooguri,
and Y. Oz, ``Large N Field Theories, String Theory and Gravity,''
hep-th/9905111.

\bibitem{fourthirds} S. S. Gubser, I. R. Klebanov, and A. W. Peet,
``Entropy and temperature of black 3-branes,'' Phys. Rev. D {\bf 54}
(1996) 3915, hep-th/9602135.

\bibitem{roberto1} A. Chamblin, R. Emparan, C. V. Johnson, and R. C. Myers,
``Charged AdS Black Holes and Catastrophic Holography,'' hep-th/9902170.

\bibitem{roberto2} A. Chamblin, R. Emparan, C. V. Johnson, and R. C. Myers,
``Holography, Thermodynamics and Fluctuations of Charged AdS Black Holes,'' 
hep-th/9904197.

\bibitem{steve} S. S. Gubser, ``Thermodynamics of spinning
D3-branes,'' Nucl. Phys. {\bf B551} (1999) 667, hep-th/9810225.

\bibitem{steve2} M. Cveti\v{c} and S. S. Gubser, ``Thermodynamic Stability
and Phases of General Spinning Branes,'' hep-th/9903132.

\bibitem{steve3} M. Cveti\v{c} and S. S. Gubser, ``Phases of R-charged
Black Holes, Spinning Branes and Strongly Coupled Gauge Theories,''
JHEP 9904 (1999) 024, hep-th/9902195.

\bibitem{hologerard}  G. 't Hooft, ``Dimensional Reduction in Quantum
Gravity,'' University of Utrecht pre-print THU-93/26, gr-qc/9310026.

\bibitem{hololenny} L. Susskind, ``The World as a Hologram,''
J. Math. Phys. {\bf{36}} (1995) 6377, hep-th/9409089.

\bibitem{hawkingpage} S. W. Hawking and D. N. Page, ``Thermodynamics
of Black Holes in Anti-de Sitter Space,'' Commun. Math. Phys. {\bf 87}
(1983) 577, hep-th/9811056.

\bibitem{holographyads} E. Witten, ``Anti de Sitter Space and
Holography,'' Adv. Theor. Math. Phys. {\bf 2} (1998) 253, hep-th/9802150.

\bibitem{5dadskerr} S. W. Hawking, C. J. Hunter, and
M. M. Taylor-Robinson,  ``Rotation and the AdS/CFT correspondence,''
Phys. Rev. D {\bf 59} (1999) 064005, hep-th/9811056.

\bibitem{thermalphase} E. Witten, ``Anti-de Sitter Space, Thermal
Phase Transition, And Confinement In Gauge Theories,''
Adv. Theor. Math. Phys. {\bf 2} (1998) 505, hep-th/9803131.

\bibitem{finnmirjam} M. Cveti\v{c} and F. Larsen, ``Near Horizon Geometry
of Rotating Black Holes in Five Dimensions,'' Nucl. Phys. {\bf B531}
(1998) 239, hep-th/9805097.

\end{thebibliography}
\end{document}